\documentclass[12pt, prd, showpacs]{revtex4}
\usepackage{amssymb}
\usepackage{amsmath}

\input{tcilatex}

\begin{document}

\title{Schwarzschild black hole as particle accelerator of spinning particles%
}
\author{O. B. Zaslavskii}
\affiliation{Department of Physics and Technology, Kharkov V.N. Karazin National
University, 4 Svoboda Square, Kharkov 61022, Ukraine}
\affiliation{Institute of Mathematics and Mechanics, Kazan Federal University, 18
Kremlyovskaya St., Kazan 420008, Russia}
\email{zaslav@ukr.net }

\begin{abstract}
It is shown that in the Schwarzschild background there exists direct
counterpart of the Ba\~{n}ados-Silk-West effect for spinning particles. This
means that if two particles collide near the black hole horizon, their
energy in the centre of mass frame can grow unbounded. In doing so, the
crucial role is played by so-called near-critical trajectories when
particle's parameters are almost fine-tuned. Direct scenario of the
collision under discussion is possible with restriction on the
energy-to-mass ratio \thinspace $E/m<\frac{1}{2\sqrt{3}}$ only. However, if
one takes into account multiple scattering, this becomes possible for $E\geq
m$ as well.
\end{abstract}

\maketitle

\section{Introduction}

Several years ago, it was noticed that collision of two particles falling
towards the Kerr extremal black hole can lead to unbounded growth of the
energy $E_{c.m.}$ in the centre of mass frame \cite{ban}. It is called the Ba%
\~{n}ados-Silk-West (BSW) effect after the names of its authors. This
interesting observation triggered a lot of works on this subject. It turned
out that the effect exists also for nonextremal black holes \cite{gp}, it is
inherent to generic rotating black holes \cite{prd}, etc. Quite recently, a
new venue appeared for the effect in question - collision of spinning
particles. It was considered in \cite{bans} for the Schwarzschild metric and
in \cite{gaos} for the Kerr one. As near the Kerr black hole high energy
collisions are possible even without spin, this looks like some
generalization of the same BSW effect. Meanwhile, the high energy collision
in the Schwarzschild background is a qualitatively new phenomenon. For
spinless particles there is no counterpart of it.

Acceleration of particles to unbounded energies in the Schwarzschild
background obtained in \cite{bans} looks very much unlike the BSW effect. In
particular, relevant collision can occur far from the horizon. However, the
problem is that these results are accompanied with serious physical
difficulties. The main points here are the unavoidable appearance of
superluminal motion and change of the character of trajectories from
timelike to spacelike. (See Sec. VII of \cite{bans} where all these
difficulties are discussed in detail.) Therefore, being formally correct,
the results of \cite{bans} leave very serious questions and doubts.

The aim of the present work is to show that the direct counterpart of the
BSW effect in the Schwarzschild background (overlooked in \cite{bans}) does
exist for spinning particles. In doing so, no difficulties with superluminal
motion appear. Therefore, although for spinless particles such an effect is
absent, for spinning ones it is safely included, with minor modifications,
into the general scheme elaborated for spinless particles.

Throughout the paper, the fundamental constants $G=c=1$.

\section{Basic formulas}

We consider the Schwarzschild metric 
\begin{equation}
ds^{2}=-fdt^{2}+\frac{dr^{2}}{f}+r^{2}(d\theta ^{2}+\sin ^{2}\theta d\phi
^{2})\text{,}
\end{equation}%
where $f=1-\frac{2M}{r}$, $M$ is the black hole mass. Let a spinning
particle move in this background. We restrict ourselves by motion within the
equatorial plane $\theta =\frac{\pi }{2}$ with the spin perpendicular to the
plane of motion. In the framework of the Lagrangian theory of the spinning
particle \cite{h} applied to the Schwarzschild metric, one can obtain the
expression for the component of the four-momentum: 
\begin{equation}
P^{t}=\frac{X}{f}\text{, }X=\frac{E-\frac{MsJ}{r^{3}}}{1-\frac{Ms^{2}}{r^{3}}%
}\text{,}  \label{t}
\end{equation}%
\begin{equation}
P^{\phi }=\frac{L_{eff}}{r^{2}}\text{, }L_{eff}=\frac{J-Es}{(1-\frac{Ms^{2}}{%
r^{3}})}\text{,}  \label{pl}
\end{equation}%
\begin{equation}
P^{r}=\sigma Z\text{,}  \label{Z}
\end{equation}%
where $\sigma =\pm 1$ depending on the direction of radial motion and%
\begin{equation}
Z=\sqrt{X^{2}-f[m^{2}+\frac{L_{eff}^{2}}{r^{2}}]}.  \label{zz}
\end{equation}

We assume that $P^{t}>0$ (this is direct counterpart of the forward-in-time
condition in case of spinless particles). Derivation of eqs. (\ref{t}) - (%
\ref{zz}) can be found in different works, e.g. \cite{ne}, \cite{98}, (see
eqs. 29 - 31 there). They coincide with eqs. (8) - (10) of \cite{bans},
where a reader can also find further relevant literature on equations of
motion for spinning particles. Here, $J$ is the total angular momentum, $E$
being the Killing energy, $s$ spin per unit mass. One can check that these
equations agree with those for the Kerr black hole if one puts $a=0$ in
corresonding equations of motion (see, e.g., eqs. 25, 26 of \cite{gaos}).

Then, one can calculate the energy in the centre of mass frame. According to
the standard definition,%
\begin{equation}
E_{c.m.}^{2}=-P_{\mu }P^{\mu }=m_{1}^{2}+m_{2}^{2}+2\alpha \text{, }
\end{equation}%
where $P^{\mu }=P_{1}^{\mu }+P_{2}^{\mu }$ is the total momentum of two
particles in the point of collision,%
\begin{equation}
\alpha =-P_{1\mu }P^{2\mu }\text{.}
\end{equation}

For spinless particles, one can identify $\alpha =m_{1}m_{2}\gamma $, where $%
\gamma =-u_{1\mu }u^{2\mu }$ has the meaning of the Lorentz factor of
relative motion. However, for spinning ones, such a simple interpretation is
not possible since, in general, $P^{\mu }\neq mu^{\mu }$, where $u^{\mu }$
is the four-velocity.

Direct calculation gives us%
\begin{equation}
\alpha =\frac{X_{1}X_{2}-Z_{1}Z_{2}}{f}-\frac{\left( L_{1}\right)
_{eff}\left( L_{2}\right) _{eff}}{r^{2}}\text{.}  \label{al}
\end{equation}

Two cases should be separated here. The potential divergences can occur if
for one of particles (say, particle 1) $1-\frac{Ms_{1}^{2}}{r_{0}^{3}}=0$ in
some point $r_{0}$. This case was analyzed in \cite{bans}. However, as is
mentioned above, this leads to a number of physical difficulties and it is
unclear how to resolve them. In the present Letter, we consider another case
when collision occurs near the horizon outside it. Therefore, we require $%
r_{0}<r_{H}=2M$. Then, the only potential origin of divergencies is
collisions near the horizon where $f$ is small.

As usual in the BSW effect, the crucial point is suitable classification of
trajectories. We call a particle usual if $X_{H}\neq 0$. ((Hereafter,
subscript "H" means that the corresponding quality is taken on the horizon $%
r=r_{H}.$) It is, by definition, critical if $X_{H}=0$. This means that for
the critical particle%
\begin{equation}
E-\frac{MsJ}{r_{H}^{3}}=0\text{.}  \label{ej}
\end{equation}

Then, near the horizon,%
\begin{equation}
X\approx \frac{3MsJ(r-r_{H})}{r_{H}^{4}(1-\frac{Ms^{2}}{r_{H}^{3}})}\text{.}
\label{xh}
\end{equation}

Thus in the point of collision $r=r_{c}$ (hereafter subscript "c"
corresponds to the point of collision) $X_{c}=O(f_{c}),$ the second terms in 
$Z^{2}$ (\ref{zz}) dominates, so the expression inside the square root
becomes negative. This means that such a particle cannot reach the horizon.

And, a particle is called near-critical if $X_{H}=O(\sqrt{r_{c}-r_{H}})$.
Correspondingly, 
\begin{equation}
E-\frac{MsJ}{r_{H}^{3}}=O(\sqrt{r_{c}-r_{H}})  \label{sm}
\end{equation}%
as well.

It is easy to see that if both particles are usual, $\gamma $ is finite, the
effect is absent. If one of particles is critical, it cannot reach the
horizon at all, so the effect is absent as well. The most interesting case
arises when particle 1 is near critical, whereas particle 2 is usual. Let%
\begin{equation}
X_{H}=a_{1}\sqrt{f_{c}}+O(f_{c})\text{,}  \label{xa}
\end{equation}%
where $a_{1}$ is some finite nonvanishing coefficient.

Then, in the point of collision $X_{c}\approx X_{H}+O(f_{c})$ and%
\begin{equation}
\alpha \approx \frac{\left( X_{2}\right) _{H}}{\sqrt{f_{c}}}\left( a_{1}-%
\sqrt{a_{1}^{2}-m_{1}^{2}-\frac{\left( L_{1H}^{2}\right) _{eff}}{r_{H}^{2}}}%
\right) ,  \label{ga}
\end{equation}%
where we neglected the last term in (\ref{al}) since it remains finite.
Taking into account (\ref{pl}), (\ref{ej}), we can rewrite (\ref{ga}) in the
form%
\begin{equation}
\alpha \approx \frac{\left( X_{2}\right) _{H}}{\sqrt{f_{c}}}\left( a_{1}-%
\sqrt{a_{1}^{2}-m_{1}^{2}-16\frac{M^{2}}{s^{2}}E_{1}^{2}}\right) .
\label{ga1}
\end{equation}

We see that (\ref{ga1}) diverges when $f_{c}\rightarrow 0$. Thus we obtained
the effect of unbounded growth of energy in the centre of mass frame. This
is the key observation of the present article.

It is worth stressing that it is participation of a near-critical (but not
exactly critical) particle which plays a crucial role. If particle 1 is
critical (this was assumed in \cite{gaos} for the Kerr background), it
cannot reach the horizon and, again, there is no effect. It is adjustment
between the deviation of $X_{H}$ from zero and proximity to the horizon that
makes the effect possible. More precisely, the validity of (\ref{xa}) with
small $f_{c}$ is required. According to (\ref{sm}), this is equivalent to
the requirement that $E-\frac{sJ}{8M^{2}}$ has the order $\sqrt{r_{c}-2M}$.
One should also ensure the positivity of the expression inside the square
root in (\ref{ga1}). Thus any trajectory of this kind with $%
a_{1}^{2}>m_{1}^{2}+16\frac{M^{2}}{s^{2}}E_{1}^{2}$ is suitable for our
purpose to reach unbounded $E_{c.m.}$.

\section{Avoidance of superluminal motion}

For spinning particles, the relation between the four-velocity $u^{\mu }$
and momentum $P^{\mu }$ is more complicated than for spinless ones.
According to eqs. (12), (13) of \cite{bans},%
\begin{equation}
\frac{u^{r}}{u^{t}}=\frac{P^{r}}{P^{t}}\text{,}  \label{urt}
\end{equation}%
\begin{equation}
\frac{u^{\phi }}{u^{t}}=\frac{(1+\frac{2Ms^{2}}{r^{3}})}{(1-\frac{Ms^{2}}{%
r^{3}})}\frac{P^{\phi }}{P^{t}}\text{.}  \label{uf}
\end{equation}

Then, direct calculations gives us eq. (22) of \cite{bans},%
\begin{equation}
\frac{u_{\mu }u^{\mu }}{\left( u^{t}\right) ^{2}}=-(1-\frac{2M}{r})^{2}\frac{%
(1-\frac{Ms^{2}}{r^{3}})^{2}}{(e-\frac{Msj}{r^{3}})^{2}}(1-\chi )\text{,}
\label{uu}
\end{equation}%
\begin{equation}
\chi =\frac{3Ms^{2}(j-es)^{2}(2+\frac{Ms^{2}}{r^{3}})}{r^{5}(1-\frac{Ms^{2}}{%
r^{3}})^{4}}\text{,}  \label{hi}
\end{equation}%
$e=\frac{E}{m}$, $j=\frac{J}{m}$.

There are potential divergences in (\ref{al}) near the point $r_{0}$ $%
=(Ms^{2})^{1/3}$. However, this leads to difficiulties connected with the
inevitable change of sign of $u_{\mu }u^{\mu }$ according to (\ref{uu}),
superluminal motion and causality problems \cite{bans}. However, we have
shown above that there is also another possibility that can lead to
unbounded $E_{c.m.}$ It is realized for collisions near the horizon. We are
interested in the outside region $r\geq 2M$ only and want to have $u_{\mu
}u^{\mu }<0$ everywhere in this region to avoid problems with superluminal
motion. This entails requirement $\chi <1$. Assuming the forward-in-time
condition \thinspace $X\geq 0$, we have from (\ref{xh}) that $%
r_{H}^{3}>Ms^{2}$, so%
\begin{equation}
\frac{s^{2}}{8M^{2}}\equiv x<1\text{,}  \label{x}
\end{equation}%
we see that $\chi $ is a monotonically decreasing function of $r$.
Therefore, for our purpose, it is sufficient to require 
\begin{equation}
\chi (2M)<1  \label{1}
\end{equation}%
since for any $r>2M$ we will have $\chi (r)<\chi (2M)<1$ as well. As we are
interested in trajectories giving unbounded $E_{c.m.}$, one of particle is
usual while the other one is near-critical according to explanations given
above. As far as a usual particle is concerned, it is sufficient to take a
spinless one, $s=0$. Then, $\chi =0$, so (\ref{1}) is satisfied trivially.
If $s\neq 0$ but is small enough, (\ref{1}) is obeyed by continuity. For
finite nonzero $s$, $\left\vert j-es\right\vert $ should be small enough
according to (\ref{hi}), (\ref{1}). This can be satisfied easily since $j$
and $e$ are independent quantities. We assume that this inequality holds
true for usual particles.

For the near-critical particles, $j$ and $e$ are related according to (\ref%
{ej}). In the first approximation, neglecting the small difference between
near-critical and critical trajectories, one obtains from (\ref{ej}), (\ref%
{x}), (\ref{1})%
\begin{equation}
e^{2}<\rho (x)\equiv \frac{(1-x)^{2}}{6(2+x)}\text{.}  \label{ro}
\end{equation}%
As $\rho (x)$ is the monotonically decreasing function of $x$, this entails $%
\rho (x)\leq \rho (0)$, whence%
\begin{equation}
e<\sqrt{\rho (0)}=\frac{1}{2\sqrt{3}}\text{.}  \label{e}
\end{equation}

According to (\ref{ej}), condition (\ref{ro}) can be rewritten in the form%
\begin{equation}
j<\frac{s(1-x)}{x\sqrt{6(2+x)}}<\frac{s}{2\sqrt{3}x}=\frac{4M^{2}}{\sqrt{3}s}%
.
\end{equation}%
Thus there are restrictions on the relation between the total and spin
momenta to avoid superluminal motion.

It follows from (\ref{ej}) that for the near-critical particle in (\ref{pl})
the quantity $L_{eff}\approx \frac{8M^{2}E}{s}>0$, so the relevant orbit
required for the unbounded $E_{c.m.}$ is prograde only.

As for a particle at flat infinity $e\geq 1$, this means that direct
scenario of high energy collision cannot be realized for particles falling
from infinity. However, it occurs for a particle that starts from the
intermediate region with $r\gtrsim 2M,$where the inequality (\ref{e}) can be
satisfied. Moreover, for a particle falling from infinity this is also
possible in the scenario of multiple scattering instead of direct collision.
This implies that a particle comes from infinity to the near-horizon region,
collides there with another particle and, having obtained near-critical
parameters as a result of such a collision, produces high $E_{c.m.}$ in the
next collision.

\section{Discussion and conclusions}

The obtained result shows close analogy between high energy collisions of
spinning and spinless particles near nonextremal black holes. As was found
by Grib and Pavlov \cite{gp} (see also generalization in \cite{prd}), if two
particles collide near the nonextremal horizon, the unbounded $E_{c.m.}$ is
possible, provided one of particles is not exactly critical but slightly
deviates from the critical trajectory. In doing so, it is necessary that
deviation from the critical relation of parameters have the same order as
the small lapse function $\sqrt{f_{c}}$ in the point of collision. Both for
spinless and spinning particles, the effect of unbounded $E_{c.m.}$ is
absent near nonextremal black holes, if a near-critical particle falls from
inifnity. But it becomes possible due to the scenario of multiple collisions
suggested in \cite{gp} for spinless particles.

Our result is solid since no troubles about causality and superluminal
motion occur. Thus the Schwarzschild black hole can indeed work as
acceleration of spinning particles.

\begin{acknowledgments}
This work was funded by the subsidy allocated to Kazan Federal University
for the state assignment in the sphere of scientific activities. It was
inspired by the lecture of J. W. van Holten on motion of spinning particles
in the Schwarzschild background during WE-Heraeus-Seminar "Relativistic
Geodesy: Foundations and Applications".
\end{acknowledgments}

\end{document}